\documentclass{osa-article}

\journal{oe}

\DeclareMathOperator*{\argmin}{argmin}

\articletype{Research Article}

\begin{document}

\title{A diffuser-based computational imaging funduscope}

\author{Yunzhe Li,\authormark{1} Gregory N. McKay,\authormark{2}  Nicholas J. Durr,\authormark{2,*} and Lei Tian\authormark{1,*}}

\address{\authormark{1}Department of Electrical and Computer Engineering, Boston University, Boston, MA 02215, USA\\
\authormark{2}Department of Biomedical Engineering, Johns Hopkins University, Baltimore, MD 21218, USA\\
\authormark{*}co-corresponding authors: ndurr@jhu.edu and leitian@bu.edu\\
}




\begin{abstract}
Poor access to eye care is a major global challenge that could be ameliorated by low-cost, portable, and easy-to-use diagnostic technologies. Diffuser-based imaging has the potential to enable inexpensive, compact optical systems that can reconstruct a focused image of an object over a range of defocus errors. Here, we present a diffuser-based computational funduscope that reconstructs important clinical features of a model eye. 
Compared to existing diffuser-imager architectures, our system features an infinite-conjugate design by relaying the ocular lens onto the diffuser.  
This offers shift-invariance across a wide field-of-view (FOV) and an invariant magnification across an extended depth range.
Experimentally, we demonstrate fundus image reconstruction over a 33$^{\circ}$ FOV and robustness to $\pm$4D refractive error using a constant point-spread-function. Combined with diffuser-based wavefront sensing, this technology could enable combined ocular aberrometry and funduscopic screening through a single diffuser sensor.
\end{abstract}

\section{Introduction}

Over a billion people worldwide currently suffer from poor vision that could be improved with prescription eyeglasses \cite{Durr2014}. A major barrier to providing eyeglasses to this population is access to the trained personnel and expensive equipment required for a comprehensive eye examination. Economic restrictions, a lack of clinical providers, and distance to healthcare settings all limit access to effective ocular diagnosis and treatment \cite{Durr2014,Resnikoff2012,Fletcher1999}. To overcome these barriers, eye care providers, such as Aravind Eye Care System and L V Prasad Eye Institute, are implementing an approach that relies heavily on point-of-care screening provided by trained residents at the community level \cite{Rao2004, Rangan2007, Rao2012}. This strategy alleviates issues with transportation and reduces cultural barriers that prevent uptake of services \cite{Rao2012}. For these minimally-trained workers to be effective, there is a need for low-cost, portable, and easy-to-use devices capable of the objective assessment of a wide variety of ophthalmic diseases.

A comprehensive eye examination includes both refraction to provide eyeglass prescriptions and inspection to screen for abnormalities. Numerous inexpensive, point-of-care tools for managing ophthalmic conditions have recently been developed. With the widespread adoption of mobile phones and rapid advancement of their camera technology, the potential of smartphone-based ophthalmic imaging has been recognized \cite{Lord2010,Giardini2014,Mamtora2018,Kim2018,Arima2019}. Other portable techniques have been demonstrated with an inexpensive point-and-shoot camera \cite{Tran2012} and with a Raspberry Pi \cite{Shen2017}. A computational single-pixel ophthalmoscope was recently demonstrated for increased visibility through opacities \cite{Lochocki2016}. 
To improve efficiency and quality of prescribing eyeglasses, accurate, low-cost ocular aberrometry has been demonstrated with both handheld \cite{Durr2015,Durr2019} and smartphone-based autorefractors \cite{Ciuffreda2015}. It is likely that many such devices will be required to tackle the diverse causes of global visual impairment such as age-related macular degeneration, glaucoma, and uncorrected refractive error \cite{Bourne2020}. A simple device that combines funduscopy and aberrometry, two essential parts of a comprehensive eye exam, could reduce costs and training time while increasing efficiency.

Diffuser cameras have been explored as an attractive alternative to conventional approaches to measure lightfield information  \cite{Antipa2016,platt2001history,ng2005light,li2019fast,levoy2006light,xue2020singleshot,chen2020design}. 
With a thorough characterization of the diffuser's caustic point-spread-function (PSF) under incoherent illumination, plenoptic analysis \cite{ng2005light} allows view synthesis and digital refocusing \cite{Antipa2016}. 
Lens-free diffuser-based imaging can also be implemented by formulating the reconstruction as an inverse problem, enabling single-shot volumetric photography \cite{Waller2018} and microscopy  \cite{kuo2020chip}.
Further, utilizing temporal multiplexing by the rolling shutter of a CMOS sensor, video reconstruction is possible from a single diffuser image \cite{antipa2019video}. 
Extended depth-of-field photography has been demonstrated by deconvolving with the invariant far-field diffuser PSF \cite{Cossairt2010}. 
Diffuser-based phase imaging under spatially partially coherent illumination \cite{LuL2019,Wang2020} is possible by solving the transport-of-intensity equation \cite{teague1983deterministic,petruccelli2013transport}.
Wavefront sensing \cite{Erto2017} and ocular aberrometry \cite{McKay2019} have also recently been demonstrated with diffusers. 
In each instance, the diffuser can be treated as a randomly distributed microlens array with slightly varying foci~\cite{Antipa2016}.
However, compared to the spot pattern formed by a microlens array, the caustic pattern formed by the diffuser is more structured that produces a better evenly distributed transfer function.  
This in turn makes the deconvolution problem better conditioned and minimizes the reconstruction artifacts~\cite{kuo2020chip}.

Diffuser imaging is enabled by computational imaging, which synergistically combines optics and algorithms~\cite{mait2018computational}. 
It belongs to the class of computational imaging architectures that reduce the overall optics complexity by computation, including coded aperture imaging~\cite{adams2017single,shin2019minimally}, lens-free holography~\cite{mcleod2016unconventional}, compound imager~\cite{tanida2001thin,xue2020singleshot}, and lightfield/integral imager~\cite{levoy2006light,stern2006three}. 
Diffuser imaging is attractive because it uses a simple and cheap optical element, does not require any special alignment between the diffuser and the image sensor~\cite{Antipa2016,McKay2019}, and achieves the imaging capability by a single-shot under both spatially coherent~\cite{McKay2019} and incoherent illumination~\cite{Antipa2016,kuo2020chip}.

In this paper, we develop and demonstrate funduscopy of a model eye with a diffuser-based computational imager. 
Our diffuser-imager design is particularly inspired by the work by Antipa, et al.~\cite{Antipa2016}. 
In~\cite{Antipa2016}, the diffuser is used in a ``finite-conjugate'' configuration, in which each object point  projects a spherical wavefront to the diffuser, similar to the standard ``single-lens'' imager.
In funduscopy, due to the presence of the ocular lens, the {\it ``infinite-conjugate''} configuration is better suited, in which each in-focus object point on the retina projects a {\it planar} wavefront to the diffuser.
In this configuration, the diffuser can be treated as the second lens in a telecentric 4F system, which provides both shift-invariance across the field-of-view (FOV) and an invariant magnification under defocus.
Conveniently, this configuration is identical to that used in the diffuser-based ocular aberrometry~\cite{McKay2019}.
Taken together, we believe this diffuser-based sensing framework has exciting potential advantages in ocular imaging, and could enable simultaneous funduscopy and ocular aberrometry on the same platform. 

The diffuser-imager is integrated with an illumination module based on an annular ring of LEDs, which provides flood-illumination across a 33$^{\circ}$ FOV.
We demonstrate the imaging capability of the proposed device by reconstructing various incoherent objects, including patterns on a self-emitting OLED screen and incoherently illuminated printed patterns through a simple model eye, as well as through a commercial model eye fundus. 
In addition, we assess image reconstruction quality, and demonstrate robustness of the reconstruction algorithm to refractive error imparted on the object or the PSF used for reconstruction. 
This work shows promise for the future development of a device that could perform funduscopy and aberrometry through a single diffuser camera.

\begin{figure}[t]
\centering\includegraphics[width = 0.9\linewidth]{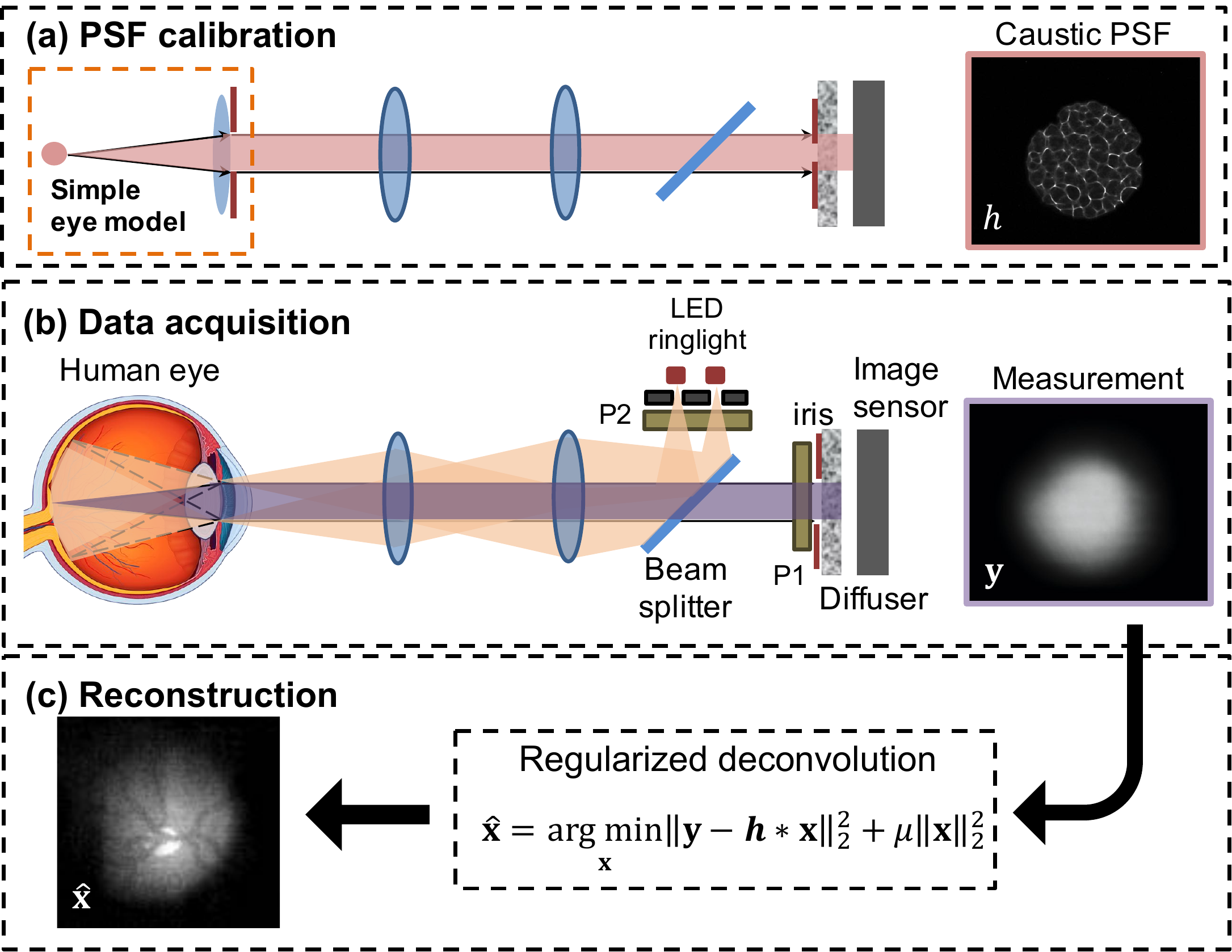}
\caption{Overview of the proposed diffuser ocular imaging workflow, including: (a) PSF calibration, (b) diffuser-image acquisition, and (c) imaging reconstruction. The PSF is a highly structured caustic pattern. During acquisition, the sensor captures a 2D image resulting from the PSF convolved with the remitted light of the fundus. A high-quality retinal image is reconstructed by solving a regularized deconvolution problem.}
\label{overview}
\end{figure}

\section{Methods}
We image the fundus with a ``DiffuserCam'', which consists of an iris placed adjacent to a holographic diffuser, separated $f_d$ from the image sensor. 
To achieve large FOV ocular imaging, we jointly designed the optical hardware, calibration procedures, acquisition methods, and reconstruction algorithms. 
Our workflow consists of three stages as summarized in Fig.~\ref{overview}, including an initial and {\it single-time} system PSF calibration, image acquisition, and computational reconstruction. 
In the PSF calibration stage, the system response is captured by displaying a point source on an OLED screen placed at the front focal plane of a simple eye model.  
In the acquisition stage, the image sensor captures a measurement of the flood-illuminated fundus through the diffuser. 
In the reconstruction stage, we solve a regularized deconvolution problem to recover a high-quality fundus image. 
In this section, we first describe our general optical design, then we lay out the specific experimental implementations for optimizing the system performance, and lastly we outline the theoretical principles and formulation of our reconstruction algorithm. 

\begin{figure}[t]
\centering\includegraphics[width = 0.9\linewidth]{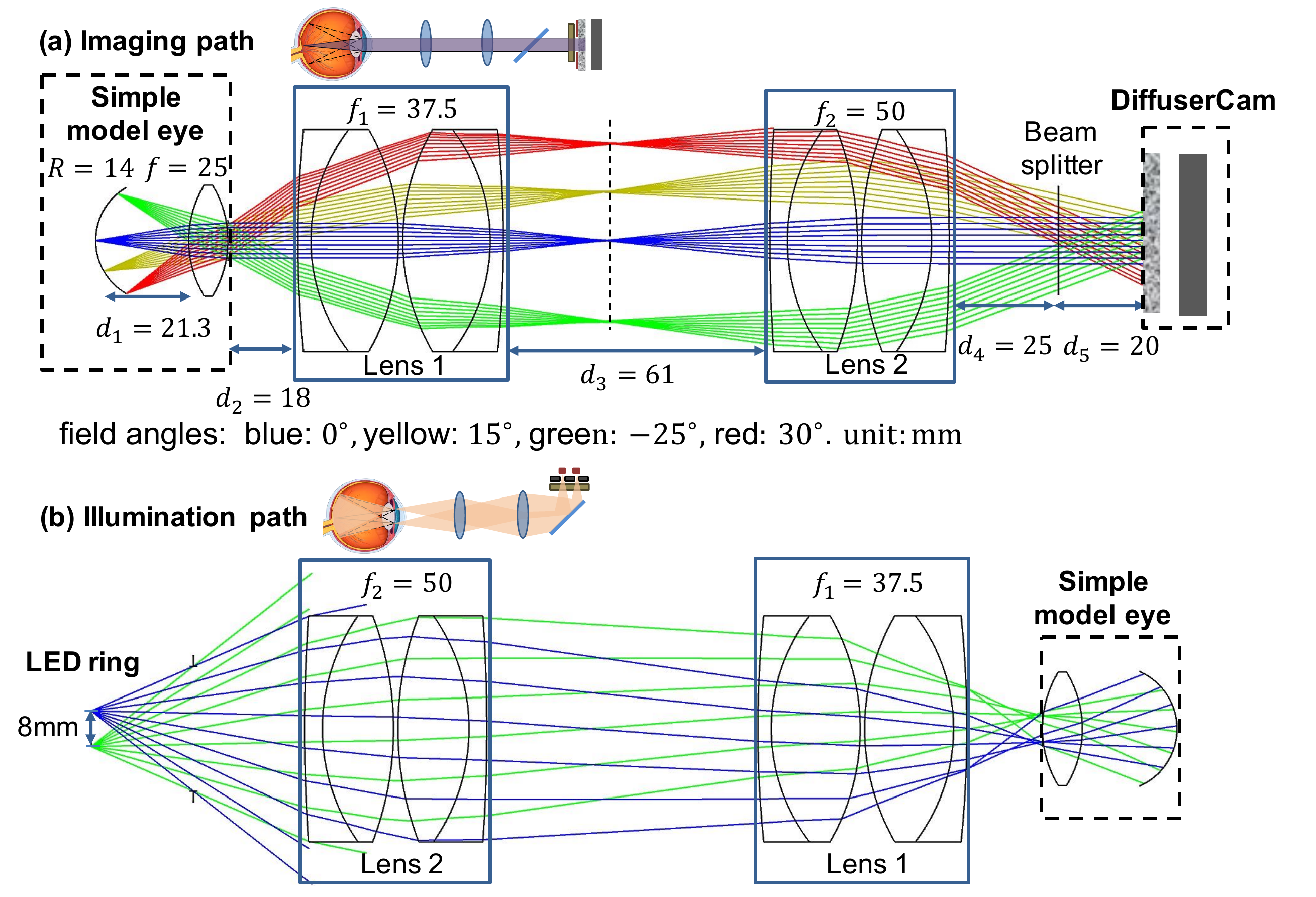}
\caption{Zemax ray-tracing of the (a) imaging and (b) illumination paths of the diffuser imager. (a) The cornea image is relayed to the diffuser so that each point on the retina produces a shifted caustic PSF. (b) Ring LEDs are also relayed to the cornea to avoid central illumination and reduce specular reflection. The FOV of the system is currently limited by the numerical aperture of the illumination optics.}
\label{method1}
\end{figure}

\subsection{Optical design}
Existing fundus cameras use lens-based imaging, where high image quality is achieved by optimizing lens design, typically resulting in multi-element, bulky, and expensive systems \cite{Shen2017}.
We propose the use of a thin diffuser as an alternative for the imaging lens, which would significantly reduce size, weight, and cost of the whole system, in addition to being compatible with diffuser-based aberrometry. 
In this system (Fig.~\ref{overview}(b)), the eye is illuminated with an LED ring through a beam splitter. 
The reflected light from the fundus first passes through a 4F relay lens system. 
The diffuser is placed at the conjugate plane of the eye lens, which captures parallel beams from an emmetropic eye. 
The image sensor is placed at the ``back-focal plane'' of a diffuser, where sharp caustic patterns form \cite{Waller2018}. 
To achieve large-FOV imaging and illumination, a carefully designed optical system is required. 
In the following, we describe our design of the imaging path and the illumination path, as shown in Fig.~\ref{method1}(a) and (b), respectively.  

To simulate an eye in a Zemax model, we use a curved retinal surface ($R$ = 14 mm), a bi-convex lens ($f = 25$ mm, Thorlabs LB1761-A), and an iris ($d = 7.7$ mm) (Fig.~\ref{method1}(a)). 
This results in collimated light exiting the pupil from any spot on the fundus. 
A relay system images the cornea to the diffuser, so that each point on the fundus produces a parallel beam with a distinct angle impinging on the same region on the diffuser within the desired FOV.
This increases the linear shift-invariance (LSI) of the system over a large FOV, which in turn simplifies the reconstruction algorithm.
To further reduce large-angle distortions, we use the Double-Gauss design \cite{mandler1980design}.
In our design, two lens groups (Lens 1 and Lens 2) each containing two doublets are inserted. 
Lens~1 has a short focal length $f_1=37.5$ mm and can collect a large angle of reflected light leaving the eye. 
The light is relayed by Lens 2 through the beam splitter (Thorlabs BSW16) and onto the diffuser. 
To avoid the beam splitter limiting the large field angle, Lens 2 should also have a short focal length $f_2=50$ mm to achieve a small magnification. 
However, the volume of the beam splitter limits the minimum length of $d_4+d_5$ to larger than 40mm. 
The relay system was designed by taking all of these constraints into consideration. 
Lens~1 consists of two $f=75$ mm doublets (Thorlabs AC508-075-A) and Lens~2 consists of two $f=100$ mm doublets (Thorlabs AC508-100-A). 
The Zemax simulation, in which we modeled the system with the actual lenses used, shows that the reflected angle within $\pm15^{\circ}$ can be well imaged to the diffuser plane without shift.
For reflected angle between $15^{\circ}$ and $25^{\circ}$, the collimated light exiting the pupil can still fully illuminate the iris adjacent to the diffuser despite the beam shift due to aberrations, indicating that the LSI condition is approximately maintained. 
For the field angle larger than $25^{\circ}$, the beam shift passes beyond the edge of the iris, which results in a different caustic PSF shape and violation of the LSI condition. For $30^{\circ}$ field angles, the shift is significant--only $6.3\%$ of the area of the entrance aperture to the DiffuserCam is illuminated by the collimated beam.

We model the thin diffuser as an array of randomly distributed microlenses with approximately the same focal length $f_d$, similar to~\cite{Antipa2016}. 
By placing the camera $f_d$ away from the diffuser, the PSF contains high-contrast caustic patterns. 
Precise control of this distance is not required because the caustic patterns from the diffuser are intrinsically more robust to defocus than a standard lens.  
Intuitively, this is because each diffuser feature can be treated as a low numerical aperture (NA) lens that provides a large depth-of-field (DOF), which removes the need for precise focus control.
Under the LSI condition, the PSF size is set by the size $d$ of the iris placed in front of the diffuser. 
The extent of the diffuser-image is approximately $d+2s$, where the maximum displacement, $s$, is related to the imaging-path FOV $\theta_{\mathrm{imag}}$ by $s = (f_1/f_2)f_d\theta_{\mathrm{imag}}/2$. 
In practice, one needs to choose an image sensor size $D > d+2s$. 
We found this condition is achievable using off-the-shelf components with a $0.5^{\circ}$ holographic diffuser (Edmunds Optics 47-989, $f_d\approx 6$ mm), the relay system described above to provide >$50^{\circ}$ FOV, which is within the range found in high-end commercial fundus cameras. 

As shown in Fig.~\ref{method1}(b), the off-axis LED ring is conjugate to eye lens through the same beam splitter and relay system. 
The achievable FOV is further limited by the illuminated area on the fundus, which approximately spans an angular FOV: $\theta_{\mathrm{illum}} = (f_2/f_1)\theta_{\mathrm{LED}}$. 
The emitting angular range of the LED $\theta_{\mathrm{LED}}$ needs to be optimized to maximize the measurement contrast. 
Our preliminary prototype uses a ring LED to minimize specular reflection with a limited $\theta_{\mathrm{LED}}$ that achieved a $33^{\circ}$ FOV.
The ring diameter was designed to be 8 mm for approximately matching the pupil size after being demagnified by the relay system, and practically constrained by the coarse LED grid in the prototype. 
By adjusting the distance between the LED ring and Lens 2, the illumination FOV can be further fine tuned.

\begin{figure}[t]
\centering\includegraphics[width = 0.75\linewidth]{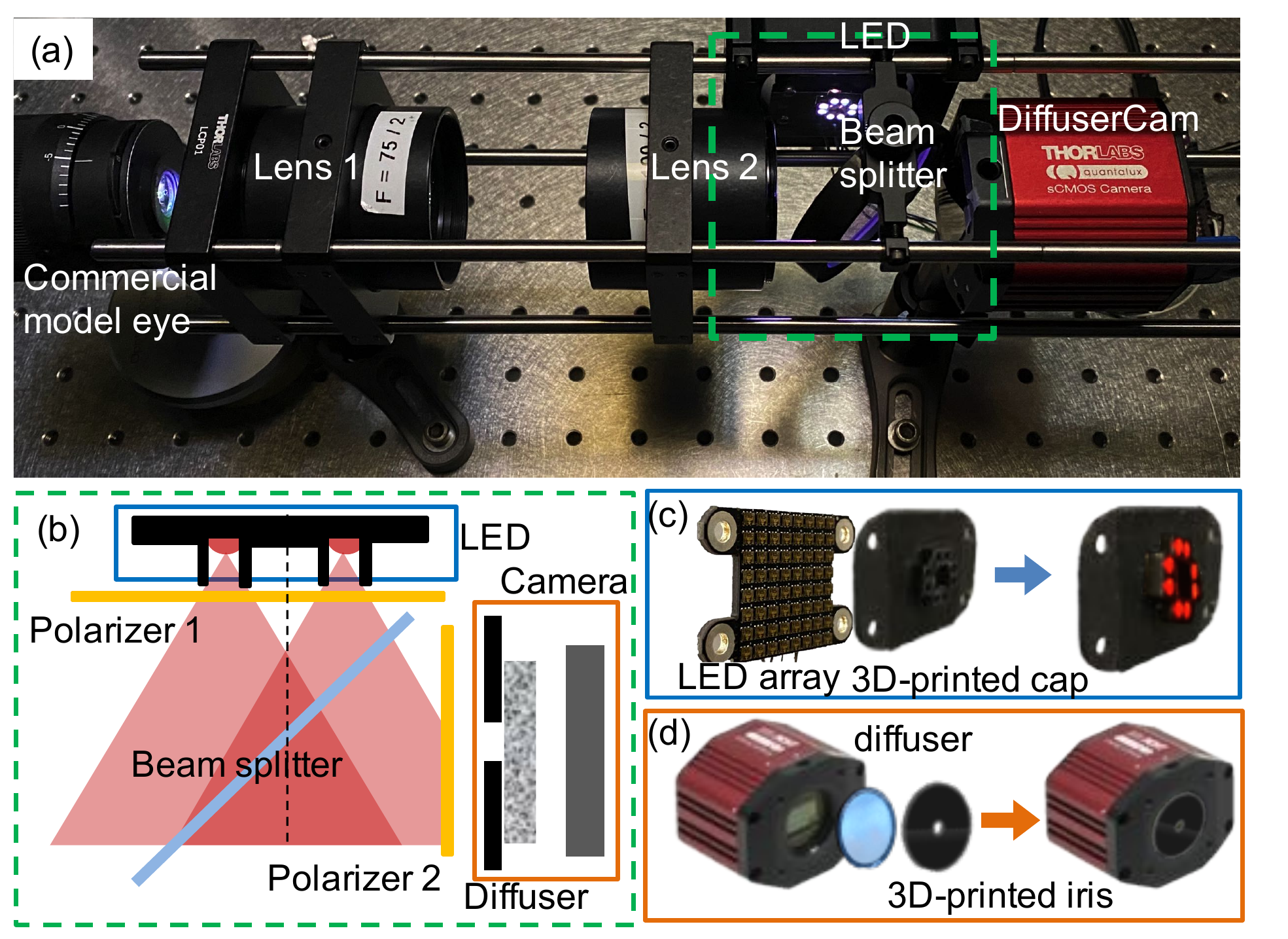}
\caption{Experimental setup of the diffuser-based funduscope. (a) Overhead view of the imaging and illumination paths, (b) details of compact components including the LED ring, a pair of polarizers crossed between the LED array and diffuser, and the diffuser camera, (c) a 3D-printed cap to confine the divergence angle of the illumination, and (d) diffuser camera components including a 3D printed iris placed in front of the diffuser to limit the size of the PSF.}
\label{method2}
\end{figure}

\subsection{Hardware design $\&$ implementation}
An overview of the diffuser-based funduscope is shown in Fig.~\ref{method2}(a). 
The setup is compact, especially for the illumination and DiffuserCam parts, as shown in Fig.~\ref{method2}(b). 
To increase the contrast and reduce the stray-light background, a pair of crossed polarizers are placed in front of the LED and the diffuser imager. 
The LED ring is implemented with an off-the-shelf LED matrix (Spacing = 2 mm, SparkFun LuMini LED Matrix - $8\times 8$ (64 $\times$ APA102-2020)). 
Due to the discrete grid of the array, the actual diameter of the ring is 9.5 mm in our prototype (Fig.~\ref{method2}(c)).
In addition, the LED array is covered by a 3D-printed cap to limit the illumination angle (Fig.~\ref{method2}(c)) in order to block the stray-light from entering the sensor directly. 
For the diffuser imager, the diffuser and a 3D-printed iris with 3.2 mm diameter are placed adjacent to one another and $\sim$6 mm before a monochromatic sCMOS image sensor (Thorlabs Quantalux, 5.04 $\mu$m pixel size, 1920 $\times$ 1080 pixels, full-well capacity 23000 e$^-$, dynamic range 87 dB).

\subsection{Algorithm}
The final fundus images are reconstructed following the general inverse problem framework that combines two complementary sources of information: the forward model describing the imaging process with the pre-calibrated PSF, and the prior describing the structural or statistical information about fundus images. 

Specifically, we used a 2D LSI model that assumes the raw measurements are the convolution of the object and a single invariant PSF that is pre-captured with an on-axis point source. 
With this LSI model, our preliminary experiments show high-resolution reconstruction in the central FOV region of a diffuser image of the retinal object. 
However, the direct deconvolution algorithm is inevitably sensitive to noise due to the poor conditioning of the inverse problem using the non-conjugate imaging geometry. We mitigated this poor-conditioning by incorporating priors in the deconvolution algorithm. The prior we used is through the L-2 norm regularizer. Accordingly, we formulate the regularized inverse problem through the minimization of: 
\begin{equation*}
\hat{\mathbf{x}}=\argmin_\mathbf{x}||\mathbf{y}-\mathbf{h}*\mathbf{x}||^{2}_{2}+\mu||\mathbf{x}||^{2}_{2},
\end{equation*}
where $\mathbf{y}$ denotes the measurement, $\mathbf{x}$ the object, $\mathbf{h}$ is the PSF, and $\mu||\mathbf{x}||^{2}_{2}$ is the L-2 regularization term with weight $\mu$. This Tikhonov regularized solution is conveniently calculated by first performing the Fourier transform, then Fourier domain filtering, and lastly inverse Fourier transforming. The optimal regularization parameter $\mu$ is found by picking the visually-optimal reconstruction when varying $\mu$ in a predefined small range.

\begin{figure}[t]
\centering\includegraphics[width = \linewidth]{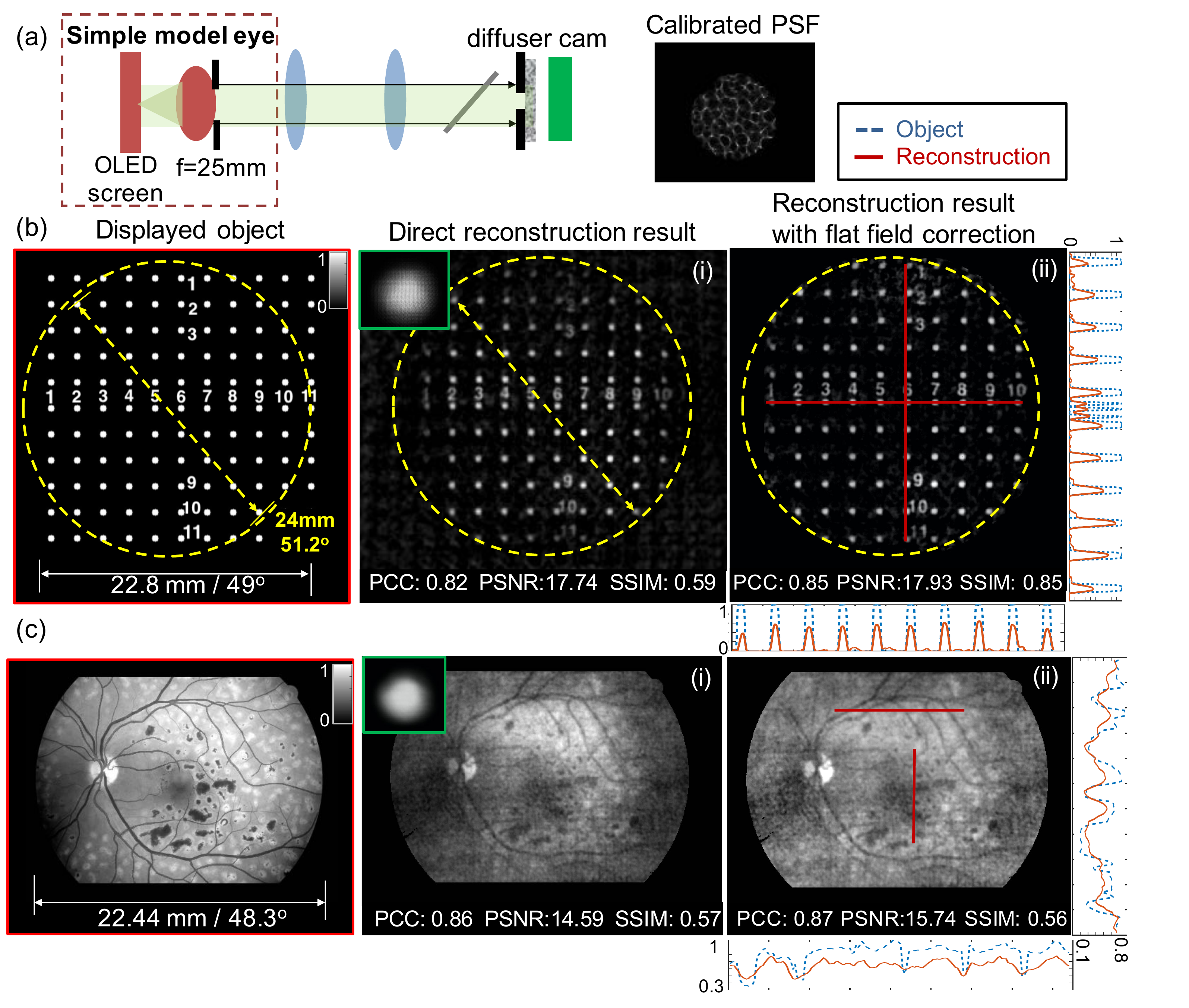}
\caption{Experimental FOV characterization of the imaging path of our prototype by using a self-illuminated retina. The design in (a) provides $\sim50^\circ$ FOV, as demonstrated on (b) a dot-array object and (c) a retinal object. The raw diffuser image acquired from each object is shown in the green-outlined inset. The direct reconstructed images in (b)(i) and (c)(i) and the ones with flat field post-correction in (b)(ii) and (c)(ii) are compared. Cutlines are compared between the flat-field corrected reconstruction and the displayed object. The PCC, SSIM, and PSNR of each reconstructed image are computed against the original displayed object.}
\label{fig:result1}
\end{figure}

\section{Results}
In this section, we assess the diffuser funduscope in three scenarios. 
First, we conducted PSF calibration and FOV analysis of the imaging system using a simple model eye with a self-illuminated OLED screen for its retina. 
Second, we replaced the self-illuminated OLED screen with various objects printed on paper, and projected the system's off-axis, ring illumination to the simple model eye for reconstruction. 
The first two experiments show the effect of the system's illumination on the FOV. 
Last, we use a commercial model eye to assess the image quality of the diffuser funduscope with a physiologically realistic object.

\subsection{Simple model eye with self-illuminating object}
\label{sec:result1}
For analyzing the FOV of the imaging system in the absence of illumination limitations, we measured a self-illuminated simple model eye (Fig.~\ref{fig:result1}(a)). 
The simple model eye is composed of an object placed at the focal plane of a biconvex lens ($f = 25$ mm) and a 7.7 mm aperture, which provides a crude model of the human eye~\cite{McKay2019}. 
We used an OLED screen as a self-illuminated object, placed  at the focal plane. 
The PSF was measured first by displaying an on-axis point source (diameter 275 $\mu$m) on the screen, which produced the caustic pattern, shown in Fig.~\ref{fig:result1}(a). 
Next, the screen displayed an array of equally spaced point sources for calibrating the FOV as shown in Fig.~\ref{fig:result1}(b). 
Here, the green outlined insert is our raw measurement and the right-hand-side is the regularized reconstruction. 
We observe in this setting that the system is able to reconstruct over a 51.2$^\circ$ angular FOV.  
Figure~\ref{fig:result1}(c) shows our reconstruction from displaying an image of a retina with diabetic retinopathy \cite{budai2013robust} on the screen, which demonstrated a relatively wide FOV of 48.3$^\circ$.
The directly reconstructed images are shown in Figs.~\ref{fig:result1}(b)(i) and (c)(i).
We observe uneven intensity distribution in these images possibly due to the mismatch between the model curved retina (in Fig.~\ref{method1}) and the flat OLED screen used for displaying the object. 
To compensate for this effect, we applied an additional flat-field post-correction to the reconstruction, as shown in Figs.~\ref{fig:result1}(b)(ii) and (c)(ii).
Cutlines are made across characteristic feature regions to compare the reconstruction and the original displayed object.

The reconstruction quality is further quantified by comparing the recovered image with the original displayed object using the Pearson correlation coefficient (PCC), peak signal-to-noise ratio (PSNR), and structural similarity index (SSIM). 
For the dot array object in Fig.~\ref{fig:result1}(b), the flat-field correction leads to reconstruction quality improvement, in particular--the SSIM improves from 0.59 to 0.85. 
This can be understood from an improvement in the luminance agreement between points at large FOV angles. For the retinal image object in Fig.~\ref{fig:result1}(c), the loss of contrast reduces the SSIM.  
Looking at the cutlines, though the vascular features are blurred in the reconstruction, the major features are captured, and both bright and dark edges are followed.

\begin{figure}[h]
\centering\includegraphics[width = \linewidth]{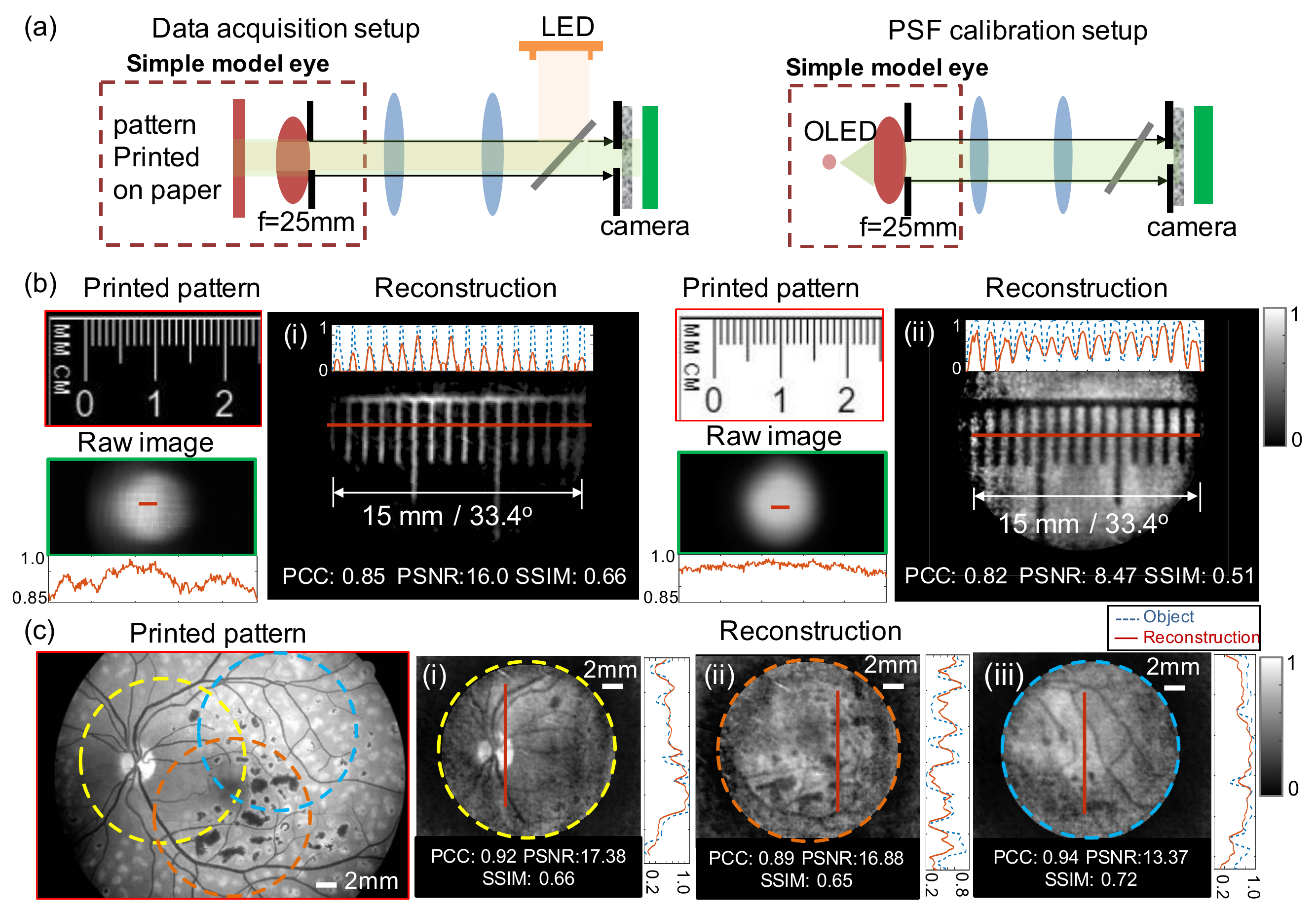}
\caption{Experimental FOV characterization of external illumination and imaging. (a)~Test objects are printed on paper and placed at the focal plane of the simple model eye. The PSF used is the same as was acquired previously from the OLED screen. (b,c) Our design provides $\sim$33$^\circ$ FOV, as demonstrated with (b) ruler patterns with both positive and negative contrast, and (c) a retinal image. All reconstructions are flat-field corrected. Cutlines of the raw images are shown to compare the measurement contrast from the positive and negative contrasted ruler patterns.
Cutlines are compared between the reconstruction and the printed pattern. The PCC, SSIM, and PSNR of each reconstructed image are computed against the original printed pattern.}
\label{fig:result2}
\end{figure}

\subsection{Simple model eye with external illumination}
Next, we tested diffuser imaging with an external illumination system. 
In this experiment, the simple model eye was also used, however we substituted the OLED screen with a printed paper object, as shown in Fig.~\ref{fig:result2}(a). 
The same PSF measured from the on-axis point source on OLED screen (from section~\ref{sec:result1}) was used for the reconstruction, as illustrated in Fig.~\ref{fig:result2}(a). 
We first analyzed the FOV by imaging a printed ruler with both positive and negative contrast (Fig.~\ref{fig:result2}(b)). 
The imaging system reconstructed an image of the ruler 15 mm in length, equivalent to a 33.4$^\circ$ angular FOV. 
We then printed and imaged the same retinal image displayed in Fig. \ref{fig:result1}(c). 
In this printed fundus, the FOV was moved to three locations by translating the model eye laterally.
Due to the planar object and uneven illumination, non-uniform intensity distribution in the direct reconstruction was also observed. 
We applied the same flat-field correction to each reconstruction in the results shown in Fig.~\ref{fig:result2}(b) and (c).
Cutlines are made across characteristic features  to compare the reconstruction and the corresponding printed pattern.
Similar to our previous observation, aside from minor reduced contrast, we show major features are faithfully recovered.
In particular, in each FOV of Fig.~\ref{fig:result2}(c), vessels and other small features can clearly be resolved.
The reconstruction quality is further inspected by PCC, PSNR and SSIM. 
As highlighted by the comparison for the ruler pattern with the positive and negative contrast in Fig.~\ref{fig:result2}(b), while the PCC remains similar, the PSNR and SSIM are much reduced for the positive contrast case (Fig.~\ref{fig:result2}(b)(ii)). 
This degradation indicates a potential challenge in reconstructing an object with sparse negative-contrasted features with an otherwise uniform bright background since it results in a less structured measurement with low contrast, as evident by comparing the cutlines from the raw measurements between the two cases in Fig.~\ref{fig:result2}(b).
For the printed retina object, the cutlines again show that important clinical features like the vessels, hemorrhages, and optic disk are captured by the reconstruction at all three FOVs explored in Fig.~\ref{fig:result2}(c).

\begin{figure}[h]
\centering\includegraphics[width = \linewidth]{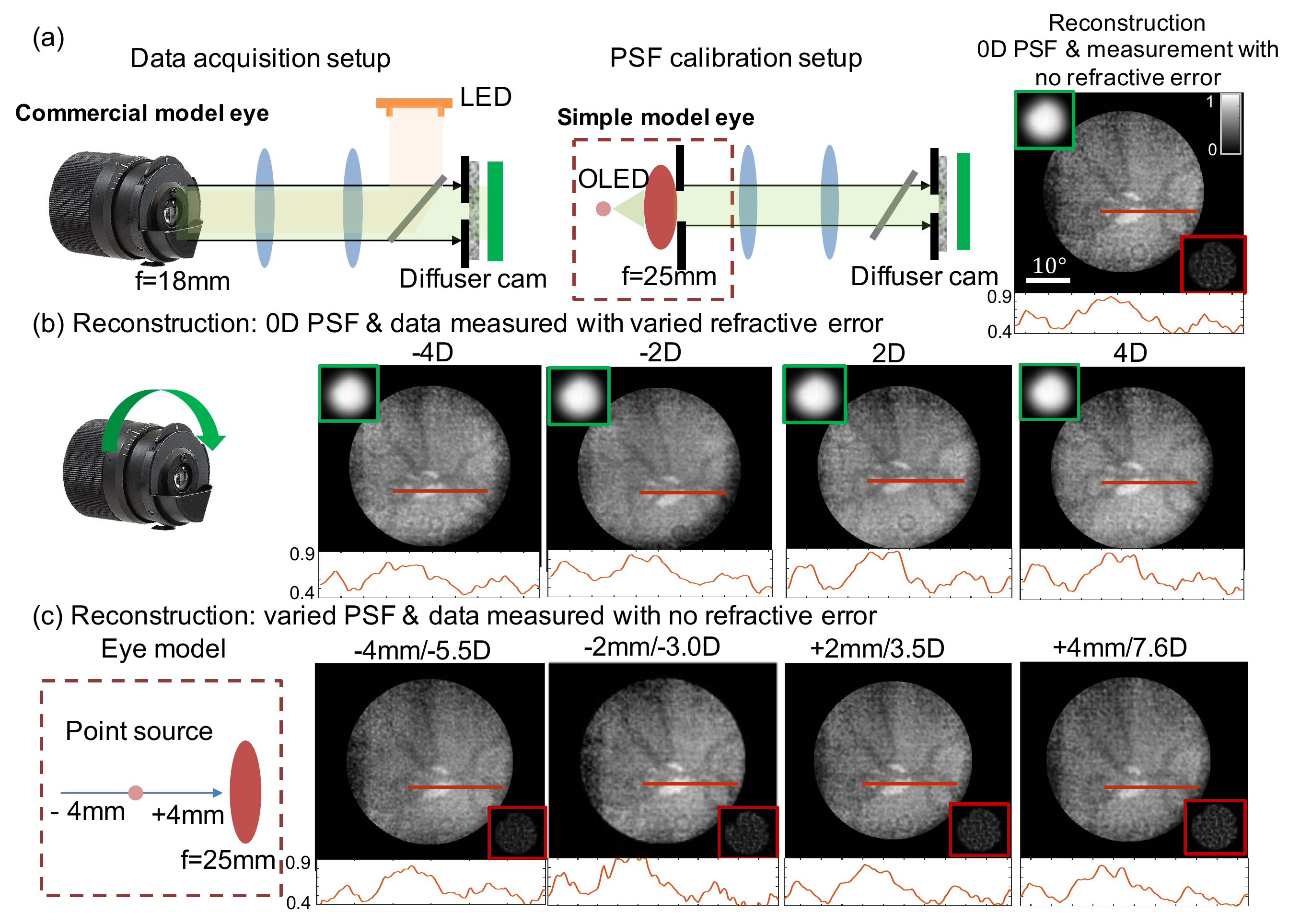}
\caption{Diffuser imaging of a commercial model eye. (a) Left: The data acquisition and PSF calibration are individually performed on different setups. Right: The reconstruction from both the measurement and the PSF taken with no refractive error (0D). (b) The reconstruction results using fundus measurements under different refractive errors with a 0D PSF. 
(c) The reconstruction results of a 0D fundus measurement using aberrated PSFs.
Cutlines are shown in each reconstructed images and demonstrate consistent contrast with different refractive errors or defocused PSFs.}
\label{fig:result3}
\end{figure}

\subsection{Commercial model eye with external illumination}
To investigate the performance of our combined imaging and illumination system in a physiologically-realistic scenario, we next imaged a commercial model eye (HEINE Ophthalmoscope Trainer), which has realistic retinal structures and allows varying amounts of refractive error. 
Figure \ref{fig:result3}(a) shows the overall procedure of this experiment. 
First, raw data is acquired (green insert) without refractive error (0D), and reconstructed using the same PSF calibrated in the previous experiments (from Fig.~\ref{fig:result1}(a)). 
Next, to assess the robustness of the system to refractive error, we tested two scenarios: (1) reconstruction of fundus images with varying refractive errors using a single emmetropic PSF acquired at 0D refractive error, and (2) reconstruction of the fundus of an emmetropic model eye using ametropic PSFs acquired at a range of different retinal positions. 
Figure~\ref{fig:result3}(b) shows that when the refractive error is within a range of -4D to 4D, no significant degradation of image quality occurs. 
In Fig.~\ref{fig:result3}(c), the distance at which the PSF was acquired was varied by $\pm$4 mm. 
This simulates PSFs acquired in eyes that are too short or too long, thus testing the sensitivity of the reconstruction to changes in the PSF that could result if calibration is done in an eye with refractive error.
Using a thin-lens approximation, this spans refractive error from -5.5D to +7.6D. 
The reconstruction results indicate that when the refractive error is as high as -5.5D or +7.6D, we observe slight degradation of image quality in both contrast and resolution. 
All reconstructed images are flat-field corrected to compensate for uneven illumination. 
A cutline across the optic disk region for each reconstructed image is shown, indicating consistent contrast is achieved across the defocus range investigated.

\section{Discussion}

\subsection{Self-illuminated object}
The initial set of experiments with a self-illuminated OLED screen used as the retina in a simple model eye provided insight into the fundamental operation and potential performance of the diffuser-based funduscope in the absence of illumination constraints. 
For initial calibration and subsequent deconvolution, the system PSF was measured by illuminating an on-axis point source on the OLED screen. 
As seen in Fig.~ref{fig:result1}(a), the PSF is a highly structured caustic pattern, which is the fundamental signal used in both DiffuserCam~\cite{Waller2018} and diffuser-based ocular aberrometry~\cite{McKay2019}.

Next, a point source array was illuminated on the screen (Fig.~\ref{fig:result1}(b)), and the acquired image (green insert) is deconvolved with the system PSF to reconstruct the object. 
From this result, we can determine that the imaging path is able to provide a 51$^{\circ}$ angular FOV. Interestingly, when looking at the acquired signal before reconstruction (Fig.~\ref{fig:result1}(b), green insert), a structured pattern is observed, consistent with the expected appearance of the PSF convolved with a point array.

When the image of a retina with diabetic retinopathy is displayed on the OLED screen (Fig.~\ref{fig:result1}(c)), a similar 48.3$^{\circ}$ FOV is reconstructed. 
Many important features of the retina are visible, including the optic disk and healthy vasculature, as well as retinal scarring and hemorrhage. 
The detection FOV is similar to the 45$^{\circ}$ FOV typically achieved by non-mydriatic fundus photography~\cite{Mackay2015}. 
The reconstructed structures at large field angles are more blurred and have lower contrast, due to distortion of a flat screen being projected by a single bi-convex lens. 
When exiting the simple model eye lens, these rays are not parallel, which changes their PSF, making it no longer spatially invariant. When applying our deconvolution algorithm with a fixed PSF, the reconstruction performance of structure at high field angle decreases.    

\subsection{External illumination with a simple model eye}
In the next set of experiments, we used external illumination via a ring LED, and demonstrated simple model eye reconstructions of test objects printed on paper (Fig.~\ref{fig:result2}). 
We first used printed rulings of known size for calibration (Fig.~\ref{fig:result2}(b)). 
From these objects, we observed a high contrast reconstruction over a 33.4$^{\circ}$ FOV. 
Comparing these results with the 51$^{\circ}$ FOV demonstrated with the OLED retina (Fig.~\ref{fig:result1}(b)), it is apparent that the current system FOV is limited by the extent of the LED illumination. 
This can be mitigated by improving the illumination numerical aperture. 
Using the same two printed ruler patterns in Fig.~\ref{fig:result2}(b) with opposite contrast demonstrates the impact of signal sparsity on measurement contrast and reconstruction quality. 
The sparse-printed rulings (white rulings and black background, left) were captured with higher contrast and reconstructed with better quality, as compared to its counterpart, the dense-printed rulings (black rulings on white background, right).

Lastly, we printed the same retina pattern used on the OLED screen (Fig.~\ref{fig:result1}(c)) for imaging and reconstruction using the ring LED illumination (Fig.~\ref{fig:result2}(c)). 
Again we observe a similar FOV as observed with the ruler, less than when we used the displayed retina on the OLED screen, further indicating our FOV is limited by the illumination extent. 
However, despite a more limited FOV, we still observe the same clinical retinal features reconstructed in the self-illuminated pattern. 

\subsection{Commercial model eye reconstruction}

In the last set of experiments, we replaced the simple model eye with a commercial model eye to test a more physiologically realistic object in the presence of refractive error and aberrated PSFs (Fig.~\ref{fig:result3}). 
Reconstructions of the commercial model eye fundus were performed with LED illumination, using the same PSF acquired from the on-axis point source in the calibration step (Fig.~\ref{fig:result1}(a)), and with 0D refractive error. 
The initial result of the commercial model eye fundus reconstruction is shown in Fig. \ref{fig:result3}(a), where the optic disk and blood vessels are clearly visible. 
A similar angular FOV was observed here with the commercial model eye as was previously demonstrated with the simple model eye. 
The FOV, though limited by the extent of illumination provided by the ring LED, still reveals a greater FOV than conventional direct ophthalmoscopy~\cite{Mackay2015}. 

After reconstructing an initial image of the commercial model eye fundus with emmetropia, a refractive error of between -4D and 4D was introduced (Fig.~\ref{fig:result3}(b)). 
Despite the refractive error, the diffuser funduscope still produced fundus images of similar quality to the emmetropic case. 
Importantly, fundus images were reconstructed using the same PSF as applied in the 0D case and in the reconstructions of Fig. \ref{fig:result2}, and no re-calibration was required.

Finally, we evaluated the image reconstruction quality produced if the PSF were acquired from locations other than the focal point of the model eye lens (Fig.~\ref{fig:result3}(c)). 
PSFs were acquired as the OLED screen was translated from -4mm to +4mm, to simulate the PSF acquired from eyes that are too long (myopia), and too short (hyperopia), respectively. 
Next, the raw data acquired from the 0D refractive error case was reconstructed with these varied PSFs. 
Again the fundus was reconstructed successfully over a similar 33$^{\circ}$ FOV, though some degradation of contrast and resolution begin to appear at the extreme refractive errors.

Together, these results demonstrate that the diffuser funduscope is robust to refractive error over a range of myopia and hyperopia large enough to cover a substantial range of clinical cases~\cite{Schwiegerling}. 
This is because the 0.5$^{\circ}$ holographic diffuser used in this study can be modeled as a random array of irregular lenslets, each with a very large f/\# and large depth-of-focus. 
Further, the reconstruction quality is similar to that achieved by other computational ophthalmoscope techniques~\cite{Lochocki2016}. 
Overall, we believe our imaging system is robust and stable to refractive error within a reasonable range, and shows promise for improving the accessibility of medical diagnosis of retinal disease.

\subsection{Special considerations in diffuser-based computational imaging}
The diffuser-imaging follows a ``non-focal'' imaging geometry -- the PSF spreads over an extended area. 
Accordingly, each pixel on the image sensor measures mixed signals from multiple object points. 
This generally reduces the image contrast in the raw measurements as compared to traditional ``focused'' imaging systems.
Consequently, image sensors with low read-noise, large well capacity, and high dynamic range are desired to better capture the encoded information. 
In our prototype, we chose an sCMOS camera that provides <1 e$^-$ median read-noise, 23000 e$^-$ full-well capacity and 87 dB dynamic range. 
Image quality can be further improved with better image sensor, as shown in~\cite{Antipa2016}.
The image contrast can also be significantly improved by using a properly-designed microlens array~\cite{xue2020singleshot} or customized aperiodic microlens array~\cite{kuo2020chip}.
Our future work will investigate these strategies with the additional considerations of keeping the platform low-cost and compatible to ocular aberrometry.

Our reconstruction was implemented by the Tikhonov regularization algorithm, which has the advantage that it provides a closed-form solution, is computationally efficient and fast.
In our implementation, reconstructing a 1080$\times$1920-pixel image using MATLAB on Mac OS machine takes about 0.26 seconds. 
However, this L-2 regularization strategy suffers from a fundamental trade-off between the reconstructed resolution, image contrast, and ringing artifacts~\cite{bertero1998introduction}. 
This limited the reconstruction quality, especially for complex  objects, e.g. Fig.~\ref{fig:result1}(c).
The reconstruction quality is further affected by the presence of a strong background, e.g. Fig.~\ref{fig:result2}(b).
It has been shown that these limitations can be alleviated by incorporating advanced image priors, such as sparsity~\cite{Antipa2016} and deep neural network learned priors~\cite{monakhova2019learned}, using an iterative algorithm.
However, these algorithms typically require significantly larger computational cost and longer execution time. 
Recently, end-to-end task-specific deep learning models emerge to be an appealing solution for achieving both high-quality image reconstruction and highly efficient inference implementation~\cite{barbastathis2019use,sinha2017lensless,li2018deep,xue2019reliable}.
Given the recent achievements in automatic analysis of retinal images~\cite{gulshan2016development,gargeya2017automated}, combining our diffuser-funduscopy and data-driven deep learning models may be a promising future direction to pursue.

\section{Conclusion}
Visual impairment is a pressing global health concern, and many of its causes are avoidable. This problem can be improved by the development and distribution of robust, low-cost diagnostic devices that require minimal training to operate. In this paper we develop and demonstrate one such approach for retinal imaging by developing and characterizing a compact diffuser funduscope. We demonstrate high-quality funduscopy of a model eye that is robust to a large range of refractive error. Further, the point spread function used for deconvolution is a caustic pattern produced by the same holographic diffuser from which ocular aberrometry has been previously demonstrated. In future work, we envision these two techniques working synergistically, allowing simultaneous measurement of refractive error and funduscopy in one compact, inexpensive device.

\section*{Funding}
This work was partially supported by a Johns Hopkins Medical Scientist Training Program Fellowship, an NIH NIBIB R21 grant (R21 EB024700), and NSF grants (1711156, 1813848).

\section*{Acknowledgments}
We thank Shivang R. Dave and Ahhyun Stephanie Nam from PlenOptika, Inc., for fruitful discussions on funduscopy. 

\section*{Disclosures}
GNM and NJD are listed as co-inventors on a provisional patent application assigned to Johns Hopkins University that is related to the technologies described in this article. They may be entitled to future royalties from this intellectual property.
\bibliography{sample}

\end{document}